\documentclass[9pt,journal,compsoc]{IEEEtran}

\usepackage{cite}
\usepackage[T1]{fontenc}
\usepackage[utf8]{inputenc}
\usepackage{booktabs}
\usepackage{tikz}
\usepackage{url}
\usepackage[hidelinks]{hyperref}
\usepackage{xcolor}
\usepackage{comment}

\newif\ifwithComments
\withCommentstrue

\usepackage{xcolor}
\usepackage{xspace}

\ifwithComments
  \newcommand{\dmg}[1]{{\color{red}dmg says: #1}\xspace}
  \newcommand{\dmgself}[1]{{\color{red}dmg note to himself: #1}\xspace}
  \newcommand{\grex}[1]{{\color{purple}grex says: #1}\xspace}
\else
  \newcommand{\dmg}[1]{}
  \newcommand{\dmgself}[1]{}
  \newcommand{\grex}[1]{}
\fi

\usepackage{framed}
\usepackage{xparse}

\newsavebox{\fminipagebox}
\NewDocumentEnvironment{hassanbox}{m O{\fboxsep}}
 {\par\kern#2\noindent\begin{lrbox}{\fminipagebox}
  \begin{minipage}{#1}\ignorespaces}
 {\end{minipage}\end{lrbox}\makebox[#1]{\kern\dimexpr-\fboxsep-\fboxrule\relax
    \fbox{\usebox{\fminipagebox}}\kern\dimexpr-\fboxsep-\fboxrule\relax
  }\par\kern#2
 }

\begin{document}

\title{Open Source Stewardship Communities: \\ ``We need you, but not your pull request''}

\author{Gregorio~Robles and
        Daniel~M.~German\IEEEcompsocitemizethanks{\IEEEcompsocthanksitem Gregorio Robles is with the Universidad Rey Juan Carlos, Madrid, Spain.\protect\\
E-mail: grex@gsyc.urjc.es
\IEEEcompsocthanksitem Daniel M. German is with the Department of Computer Science, University of Victoria, Victoria, BC, Canada.\protect\\
E-mail: dmg@uvic.ca}}


\IEEEtitleabstractindextext{\begin{abstract}

Human-centric AI for software engineering means keeping humans responsible for work performed with AI. In Open
Source Software (OSS), AI lowers the cost of implementing changes, but reviewing someone else's contribution
remains comparatively expensive, so some projects now restrict who may contribute implementations while still
welcoming other participation---not because the code is AI-generated, but because it no longer justifies the
review cost. We call the resulting form a
\textbf{stewardship community}: a small core retains implementation authority while a broader community continues
to shape the software without writing code, and access to coding increasingly depends on approval rather than
self-initiated contribution. This raises a broader question: what happens to the human community when coding agents let
maintainers replace implementation work once supplied by external contributors? For human-centric software engineering, keeping humans in control of AI agents is not
enough: AI can replace implementation labor while weakening how OSS communities renew themselves.

\end{abstract}

}

\maketitle
\IEEEdisplaynontitleabstractindextext
\IEEEpeerreviewmaketitle

\section{Introduction}

Open Source Software (OSS) guarantees freedoms over the software itself: users may inspect, modify, and redistribute the source code under the terms of its license. \textbf{It does not guarantee a right to participate in the project, to submit changes, or to have those changes reviewed or accepted.} The distinction is important because licensing issues and contribution mechanisms function as distinct axes within an OSS project.

Many OSS projects have nevertheless developed around a highly permeable model of participation, in which users can progress from reporters to contributors to maintainers~\cite{crowston-2005-social-struc,trinkenreich-2020-hidden-figur,turzo-2025-from-first}. The GitHub-driven workflow, via its pull request model, serves as the quintessential paradigm for this approach to building OSS: anyone can fork a repository (which will be called the \emph{upstream} repository), modify the source code, and request that the upstream maintainers incorporate those changes into the upstream repository through a pull request that they can review, comment on, request revisions for, or decline. Contributions are therefore more than a mechanism for obtaining code: through repeated contribution and review, participants acquire project-specific knowledge, demonstrate competence and commitment, and may gradually assume responsibility for the project. The contribution process consequently produces both software and people capable of sustaining that software.

Generative and agentic AI are beginning to alter the conditions under which this model developed~\cite{feng2026charting,zhang-2026-augmentation-dilution}. AI substantially reduces the effort required to produce an implementation, but receiving an external contribution still requires maintainers to understand its intent, evaluate its correctness and fit, integrate it with the rest of the system, and accept responsibility for maintaining it. The cost of writing code is falling faster than the cost of deciding whether it should be accepted into the project. In practice, this means pull request volume rises while review capacity stays flat, producing a bottleneck. Recent OSS contribution policies already reflect this asymmetry: projects increasingly protect scarce review capacity, require contributors to demonstrate understanding, preserve human-to-human review interactions, and reserve some newcomer-oriented work as learning opportunities~\cite{hora2026aipolicy}.

Contemporary software engineering research is increasingly tackling the AI-driven contribution bottleneck by deploying agents designed to scale up project review capacity. This evolution points to a development model where autonomous agents are used on both sides of the contribution pipeline: contribution and review. While this may reduce the technical burden of integration, whether it preserves the human and community functions of contribution and review remains unclear.

Some projects are taking a different strategy: \textbf{they are closing the external code-contribution path itself}. Their maintainers make extensive use of AI while rejecting external implementations altogether, regardless of whether those implementations were written by humans or with AI assistance. They continue, however, to invite other forms of participation, such as bug reports, requirements, documentation, and discussions. Thus, maintainers and their own agents implement changes to the codebase using knowledge supplied by the broader community. Code implementation therefore becomes concentrated within the project's development core. Restricting implementation to a small group is not itself new in OSS, but agentic development allows maintainers to produce more of the code themselves, reducing their dependence on external code contributions and making this model more attractive. That this response is becoming more widespread is also reflected in new GitHub features\footnote{``New Repository Settings for Configuring Pull Request Access''. https://github.blog/changelog/2026-02-13-new-repository-settings-for-configuring-pull-request-access/. Accessed: Aug.\ 29, 2026.}. In 2026, GitHub introduced repository-level controls to disable pull requests entirely or restrict their creation to collaborators, and later added persistent limits on pull requests from users without write access, with exemptions for trusted contributors. GitHub cites the rise of AI-assisted pull requests as a reason for introducing these controls.

This article explores the open questions and challenges this shift presents for community participation, maintenance, and the long-term sustainability of OSS communities. We analyze projects that are closing external code-contribution paths and ask what this means for the pipeline through which users become contributors and contributors become maintainers. For human-centric software engineering, the open question is not only whether maintainers retain oversight of their AI agents, but whether the human community around a project can still renew itself once that oversight no longer depends on receiving contributions from outsiders.

 \section{Stewardship Community: Gating who can code}
\label{sec:stewardship}

As already established, the permeable boundary around OSS maintainership rests on repeated contribution and review~\cite{crowston-2005-social-struc,trinkenreich-2020-hidden-figur,turzo-2025-from-first}. Code review does more than validate a patch: the interaction around
it lets contributors learn project conventions and architecture, and gives maintainers a chance to assess
their technical judgment, responsiveness, and fit within the project. Our previous
analysis of AI contribution policies shows that projects already recognize these community functions of review: some
protect good-first issues as learning opportunities, require contributors to engage directly in review discussions, and
describe review as an investment in developing future contributors and maintainers \cite{our-icse-un-published}. Review
effort therefore returns value in several ways: it incorporates useful code, maintains oversight of the technical direction of the project, and helps participants progress toward maintainership.

Agentic development changes this balance: once a maintainer understands a problem, has the relevant project context,
and can direct a coding agent, producing an implementation may cost substantially less than evaluating someone
else's~\cite{our-icse-un-published}. As maintainers can increasingly implement changes themselves faster than they can
review an equivalent external contribution, projects may have less reason to rely on external pull requests for implementation. External participation can instead remain valuable through identifying problems, explaining their context and importance, and describing the software behavior the community needs.

\begin{table}[htb]
  \caption{Examples of projects placing new restrictions on external implementation while retaining other forms of community participation.}
  \label{tab:gated-projects}
  \centering
  \footnotesize
  \setlength{\tabcolsep}{3pt}
  \begin{tabular}{p{\dimexpr0.19\linewidth-2\tabcolsep\relax}p{\dimexpr0.28\linewidth-2\tabcolsep\relax}p{\dimexpr0.55\linewidth-2\tabcolsep\relax}}
    \toprule
    \textbf{Project} & \textbf{External code} & \textbf{External participation retained} \\
    \midrule
    Codex & Invitation only & Reports, analysis, problem context \\
    Ladybird & Public PRs not accepted & Reports, testing, reductions, technical discussion \\
    Swamp & Restricted to internal team & Issues, problems, requirements \\
    Symfony Lang. Tools & PRs disabled & Issues and problem context \\
    Ghostty & Vouched contributors only & Issues, discussions, vouch requests \\
    \bottomrule
  \end{tabular}
\end{table}

Restrictions on external implementation are already appearing in OSS projects. In our analysis of AI-related contribution policies from 209 OSS projects, we identified several examples of this shift. The restrictions take different forms: some projects close
public pull requests, some permit external implementation only by invitation, and others require contributors to be
approved before they may submit implementations. Broader forms of participation remain open across these structures.
Such restrictions do not all have the same motivation. Some projects use them as part of a rejection of AI-mediated contributions. Our interest here is in a different subset: projects whose maintainers use agentic AI while restricting external implementation.
Table~\ref{tab:gated-projects} presents five projects that exhibit this emerging pattern.

OpenAI Codex illustrates this transition clearly: it shifted from accepting external contributions, which required strict verification of code origin and legal responsibility, to an invitation-only model where outsiders are encouraged to provide reports, analysis, and context rather than code implementations\footnote{https://github.com/openai/codex/blob/main/docs/contributing.md. Accessed: Aug.\ 29, 2026}. Ladybird has stopped accepting public pull
requests and continues to welcome bug reports, testing, reductions, standards discussion, security reports, and other
forms of participation; its maintainers retain responsibility for implementation and make extensive use of coding
agents \cite{ladybird-post}. Swamp similarly restricts code contribution to its internal development team and invites
outsiders to report problems for its engineers and AI agents to implement\footnote{https://stack72.dev/the-community-pull-request-is-dead/. Accessed: Aug.\ 29, 2026}. Symfony Language Tools
provides an issue-first variant: its pull request interface is disabled while issues remain open, with
maintainers using coding agents to turn sufficiently understood problems into implementations\footnote{https://symfony.com/blog/experimenting-with-issue-first-open-source-contributions. Accessed: Sep.\ 3, 2026}.

Ghostty introduces a different form of gate. First-time contributors must be vouched for by a maintainer before they
may submit pull requests; once vouched, they may submit subsequent pull requests without seeking approval for each
contribution\footnote{https://github.com/ghostty-org/ghostty/blob/main/CONTRIBUTING.md. Accessed: Sep.\ 3, 2026}. This creates an intermediate group between maintainers and the outside
community: maintainers, approved contributors, and participants who have not crossed the implementation gate. Codex also
uses invitation to control external implementation, although its public contribution policy does not make clear whether
an invitation establishes a persistent contributor status or applies only to a particular contribution.

Once software is released under an OSS license, it becomes a shared resource that others are entitled to use, study, modify, and redistribute. Maintainers may control its canonical repository and decide which changes become part of the project, but they do not exercise exclusive control over the software itself. In this sense, whether or not they explicitly conceive of their role this way, maintainers become stewards of a resource held in common.

The projects we examine make this distinction particularly visible. They retain communities around the software while concentrating authority over who may contribute implementations. The broader community can continue to identify problems, articulate needs, provide knowledge, and participate in discussion, while a smaller group assumes responsibility for implementing and incorporating changes. We call this organizational form a \textbf{stewardship community}.

We borrow the term \emph{stewardship} by analogy from institutions that manage shared resources on behalf of a community rather than for private ownership. A conservation or community land trust, for example, assumes long-term responsibility for a resource while the broader community continues to use, monitor, and care for it without controlling it directly. The relevant parallel is the separation between authority over a shared resource and participation in the community around it. This separation reflects a broader pattern in Ostrom's work on the governance of shared resources: sustainable stewardship depends on institutions that define who has decision-making authority while preserving meaningful community participation and a stake in the resource's future \cite{ostrom-1990-governing-commons}.

An OSS stewardship community follows the same logic: \textbf{a small core holds effective title over the codebase---the
authority to decide what may be implemented and merged---while a broader community continues to shape the resource
through reports, requirements, and discussion, without acquiring title through the act of code contribution itself}. We
describe entry into implementation as \emph{\textbf{gated}} in these communities, since it depends on crossing a
boundary controlled by those already inside the project rather than on submitting a pull request. The gate may
exclude unsolicited external implementation contributions entirely, admit selected contributors into the core group,
or pre-approve specific contributors or specific contributions; in every case, the outcome is the same custodial
structure---title stays with the core, participation continues without it.

Notably, land-trust stewardship is designed for perpetuity: it treats the
deliberate cultivation of future stewards as part of the custodial responsibility itself, not as a byproduct of
granting access. We argue that this is of paramount importance as well for the stewardship of OSS communities.
An OSS stewardship community risks assuming that contributor cultivation occurs naturally.
This is because \emph{gating} changes the mechanism of role mobility. As already noted, under a permeable contribution model, outsiders initiate progression toward
the core by gradually assuming
greater responsibility. Under a gated model, entry into the implementation pathway first depends on approval or
invitation from those already inside the boundary, which moves selection earlier in the process rather than eliminating
it. Once admitted, contributors may still progress through repeated contribution and increasing responsibility, as
Ghostty's vouched-contributor model illustrates; other gated structures restrict that pathway more sharply. 
The organizational change is therefore that participation alone no longer provides an open pathway into implementation and,
from there, toward maintainership and stewardship. The long-term sustainability of these models depends on how projects
develop future maintainers when entry into the contribution pathway itself is increasingly controlled by existing
project members.

 \section{Open Questions of a Stewardship Community}
\label{sec:consequences}

Because this shift toward stewardship-based OSS models is in its infancy, its long-term consequences remain to be seen. However, several open questions are already coming into focus for the near future.

\subsection{Developer diversity and responsiveness to diverse community needs}

Diversity within OSS communities has been linked to greater contributor engagement and project success, in part
because users with heterogeneous needs often modify software directly to address them \cite{daniel-2013-diversity,franke-2003-heterogeneous}. Open contribution's permeable boundary let that diversity reach the software: maintainers
retained authority over what became part of the project, but anyone could attempt an implementation and, through
repeated contribution, build the standing to eventually shape the project's direction.

With agentic development, this trade-off becomes much costlier for maintainers to uphold. As discussed above,
implementation is now cheap for maintainers to produce themselves, while triaging, reviewing, integrating, and
maintaining someone else's contribution remains comparatively costly. Gating becomes attractive precisely because it
lets maintainers redirect that scarce review effort toward priorities they set themselves, rather than absorbing
obligations they would not otherwise take on.

Open contribution gave users a path to implement needs that were not priorities of the existing development core and have those changes integrated into the project. When an outsider supplied the implementation, maintainers had to decide whether the contribution justified the review effort and whether the contributor was likely to remain involved, accepting the risk that the contribution's future maintenance might eventually fall to them. Gating changes this trade-off: users may still report needs, but their implementation depends on whether the existing core chooses to devote its attention and agent capacity to them.
As a result, users can still propose needs outside maintainer priorities, but they can no longer increase the likelihood of their adoption by providing the implementation themselves.

\textbf{Gating therefore changes both who can shape the software through implementation and which community needs are likely to become part of it.}

This shift in when selection happens introduces a specific bias: because people tend to form relationships with others similar to themselves---a well-established pattern in social networks called homophily \cite{mcpherson-2001-homophily}---insider-mediated admission may reproduce existing social networks in the contributor population and, because future maintainers emerge from that population, in the maintainer core as well. Ghostty's one-time vouch requirement illustrates this mechanism concretely: an outsider must receive endorsement from a maintainer before submitting pull requests; once vouched, the person joins a persistent group of approved contributors.

\subsection{Maintainership succession}

Agentic development can make its own long-term cost easy to miss. A
small maintainer core supported by effective agents may sustain, or even increase, substantial development and
maintenance activity while project knowledge and decision authority remain concentrated among very few people, in some
cases a single maintainer. \textbf{The project may appear sustainable in the short term even as development capacity
grows without a corresponding increase in succession capacity.}

\begin{table*}[htb]
\caption{Issues raised by gating external code contributions.}
\label{tab:gating-consequences}
\centering
\footnotesize
\begin{tabular}{p{0.22\linewidth}p{0.36\linewidth}p{0.32\linewidth}}
\toprule
\textbf{Issue} & \textbf{Effect of gating} & \textbf{Long-term concern} \\
\midrule
Contributor diversity &
Admission to coding depends more heavily on recognition, endorsement, or invitation by existing members. &
Social selection and homophily may narrow the size and diversity of the group that contributes to development. \\
Diversity of requirements &
External needs require the existing core to allocate attention and development capacity before they become part of the software. &
Needs outside maintainer priorities may be less likely to become implemented features. \\
  Maintainer succession &
                          Self-initiated code contribution no longer provides a route into the development group. &
                                                                                                                A narrower or more selectively admitted pool of future maintainers may increase bus-factor risk when key maintainers leave. \\
\bottomrule
\end{tabular}
\end{table*}

\par
\addvspace{1.5\baselineskip}
The long-term consequences of gating are therefore broader than a reduction in external pull requests. Gating changes who enters the development group, which externally generated needs become part of the software, and how future maintainers emerge. Table~\ref{tab:gating-consequences} summarizes these issues and their longer-term implications.
\textbf{The long-term viability of OSS stewardship projects will depend on the continued evolution of the software, the renewal of the people responsible for sustaining it, and the ability of the broader community that depends on it to shape its future.}

 \section{Conclusions}

The shift towards closing external code contributions is not primarily a reaction to AI-generated code. It reflects a reorganization of who may contribute implementation work: when maintainers can direct agents to implement a change more cheaply than they can review an equivalent external contribution, outside implementations become a less attractive source of development capacity, even though their slower, harder-to-observe benefits---new knowledge, perspectives, and future maintainers---remain as important as before. \textbf{AI can substitute for implementation labor without substituting for succession or for the diversity that open contribution brought to both the software and its future stewards.}

This matters specifically for human-centric software engineering because the risk is easy to miss from inside a single project: a maintainer who directs agents carefully, reviews their output, and retains final authority over what ships is doing everything human-centric AI asks of them, even as the broader community loses its direct pathway into implementation and, from there, into the core. \textbf{Keeping humans in the loop with AI, however, does not ensure that humans remain in the loop with one another.} Human oversight and community renewal are different achievements, and a project may succeed at the first while gradually losing capacity for the second. Gating therefore makes succession and diversity explicit governance concerns, not side effects to be managed later.

We do not suggest reopening unrestricted external pull requests: doing so would restore the review and integration costs that made gating attractive in the first place. The harder task is managing what is lost when that pathway closes, deliberately rather than assuming it will resolve itself. These consequences may take years to become visible, because succession and the community's long-term capacity to sustain the software unfold far more slowly than the immediate productivity gains of agentic development.

\bibliographystyle{IEEEtran}

\begin{IEEEbiography}[{\includegraphics[width=1in,height=1.25in,clip,keepaspectratio]{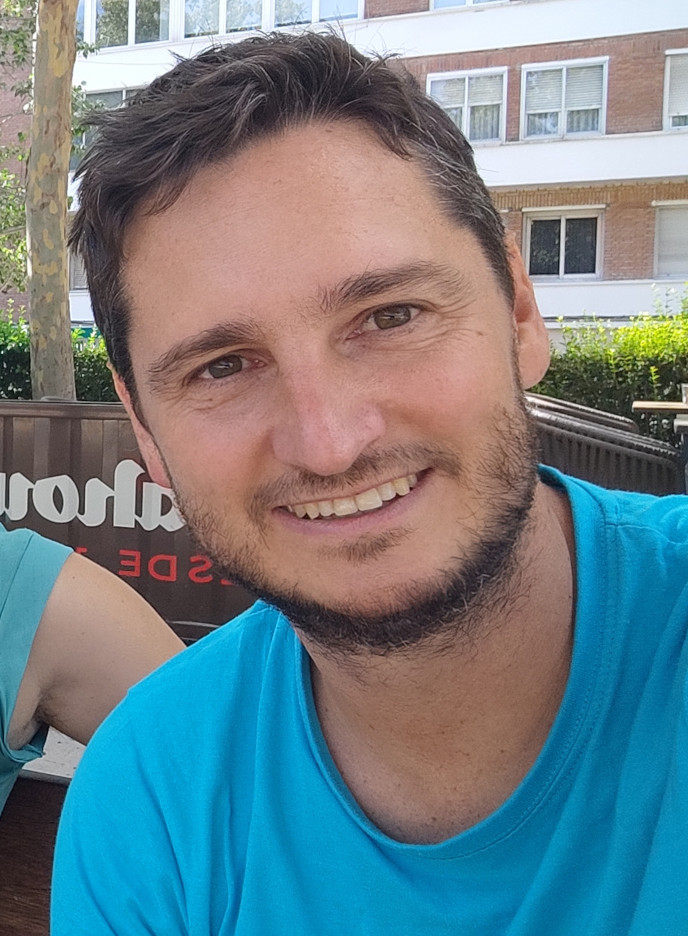}}]{Gregorio Robles} is a Full Professor at Universidad Rey Juan Carlos (URJC), Madrid, Spain. His research focuses on empirical software engineering and free/libre/open source software. Contact him at grex@gsyc.urjc.es.
\end{IEEEbiography}

\begin{IEEEbiography}[{\includegraphics[width=1in,height=1.25in,clip,keepaspectratio]{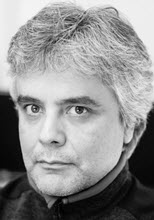}}]{Daniel M. Germán} is a Professor of Computer Science at the University of Victoria, Victoria, BC, Canada. He received his PhD from the University of Waterloo. His research focuses on open source software, software evolution, mining software repositories, and intellectual property and licensing in software development. Contact him at dmg@uvic.ca.
\end{IEEEbiography}

\end{document}